# Cascade Models in Simulation of Extended Heavy Targets Irradiated by Accelerated Proton and Deuteron Beams[a]

M.Baznat[1,2], A.Baldin[1,3], E.Baldina[1,3,b], M.Paraipan[1,4], V. Pronskikh[5,c], P.Zhivkov[6]

[1]*Joint Institute for Nuclear Research, Dubna, Russia*
[2]*Institute of Applied Physics, Chisinau, Moldova*
[3]*Institute for Advanced Studies "OMEGA", Dubna, Russia*
[4]*Institute of Space Science, Magurele, Ilfov, Romania*
[5]*Fermi National Accelerator Laboratory, Batavia, IL 60510, USA*
[6]*Institute of Nuclear Research and Nuclear Investigations, Bulgaria AS, Sofia*

Abstract

The paper presents a survey of the main numerical models used for simulation of interaction of accelerated particle beams with target nuclei. These models form the core of the software for simulation of various experiments and experimental facilities both for scientific and applied purposes. The beam and target parameters considered in detail in this study (protons and deuterons with energies from 0.66 to 4 AGeV and bulk U targets) cover the range of interest in development of new concepts of nuclear power production aided by accelerated particle beams.

INTRODUCTION

Numerical study of ADS systems, namely, interaction of accelerated proton and light ion beams with extended targets, encounters a well known problem of description of neutron breeding in such systems. Various approaches and numerical models differ in their predictions of neutron yield and spectrum up to a factor of two. That is why the comparative examination of the widely used models and codes for description of interaction of accelerated ion beams with heavy bulk targets with the idea of further experimental verification is important.

The intranuclear cascade(INC) approach was apparently first developed by Goldberger [1], who in turn based his work on the ideas of Heisenberg and Serber [2], who regarded intranuclear cascades as a series of independent collisions using on-mass-shell free particle-nucleon cross sections. The colliding particles are treated as classical point-like objects moving between collisions on well defined trajectories in the target potential well.

Many intra-nuclear cascade models have been proposed and developed in the past by several groups. In many cases the motivation was to provide a satisfactory level of description of final-state hadron spectra in the problem of few MeV to few GeV reactions of hadrons with nuclei. They find application in low-energy calorimetry, studies of nucleon shielding, accelerator based nuclear-waste degradation, neutrino beams, or studies of design and application of spallation neutron sources.

Let's remind the main basic assumptions of the INC. The main condition for the applicability of the intranuclear cascade model is that the DeBroglie wavelength $\lambda$ of the particles participating in the interaction be sufficiently small: It is necessary that for most of these particles $\lambda$ be less than the average distance between the intranuclear nucleons $\Delta \sim 10^{-13}$ cm. Only in this case does the particle acquire quasi-classical features and can we speak approximately of particle trajectory and two-particle collisions inside the nucleus. It is clear that for this to be the case the primary particle kinetic energy $T$ must be greater than several tens of MeV.

Another important condition for applicability of the INC is the requirement that the time in which an individual two-particle intranuclear collision occurs on the average, $\tau \sim 10^{-23}$ s, be less than the time interval between two such consecutive interactions

$$\Delta t = l/c \gtrsim 4\pi R^3 / 3A\sigma c \gtrsim 3 \cdot \frac{10^{-22}}{\sigma} \, (mbar)s,$$

where $l$ is the mean range of the cascade particle before the interaction, $c$ is the velocity of light, $R = r_0 A^{1/3}$ is the mean radius of the nucleus, and $\sigma$ is the cross section for interaction with an intranuclear nucleon. This permits the interaction of the incident particle with the nucleus to be reduced to a set of individual statistically independent intranuclear collisions.

Since the energy of the particles participating in the cascade is greater than the binding energy of the intranuclear nucleons – the same characteristics can be used for interaction of cascade particles inside the nucleus as for the interaction of free particles. The effect of other intranuclear nucleons is taken into account by introduction of some average potential V , and also by the action of the Pauli principle. The nucleus is considered to be a degenerate Fermi gas of nucleons enclosed in the nuclear volume. Both projectile and nucleus have to be initialized. Regarding the projectile, type and energy are known, but its impact parameter is taken randomly and coulomb deviation considered. If the projectile is a composite particle, its structure must be given in the same way as for the target. The target nucleus is defined by its mass, its charge, the potentials felt by the particles, the momentum of each nucleon (most of the time a Fermi gas distribution is used), and the spatial distribution of the nucleons. Two ways exist to define this distribution. The distribution is either continuous, often several concentric density regions, or discrete, i.e.

positions are sampled in a Wood-Saxon distribution, for example. According to the Pauli principle, the nucleons, after an intranuclear collision, must have energy above the Fermi energy; otherwise such an interaction is forbidden. The effect of Pauli principle is very important. The action of the Pauli principle leads to an increase of the mean free path of fast particles inside the nucleus. It is especially pronounced at $E_{inc}$ < 40 MeV causing to rise even though the nucleon-nucleon cross section is strongly increasing. Understanding of the limitations of INC at low energies is important for evaluation of reliability of transport calculations used in wide variety of applications. In collisions of high energy particle with the Fermi sea, the momentum transfer is small, and Pauli principle limits the interaction to small fraction of the Fermi sea close to its surface, thus increasing the mean free path. Most of the collisions are not central. Calculations show that in the energies of few tens to few hundreds MeV about 60% of the collisions leading to inelastic reactions occur at impact parameters at which the nuclear density is less than a half of the central density. The target periphery is modelled in all the INC implementations, but each has a different way to deal with the low energy participants chosen considering agreement with the experimental data rather than from basic physical considerations. The incident particle and target constituents are moving on classical trajectories in the potential well and scatter whenever their relative distance is less than $\sqrt{\sigma(E_{cm})/\pi}$, $\sigma(E_{cm})$ being the free space cross section, and $E_{cm}$ their center of mass energy.

Two different methods are applied to move and follow the particles participating to the cascade. With the time-like transport all particles are followed at the same time, while with the space-like transport the particles are followed one after another.

Three events exist during the intranuclear cascade: collision, resonance decay and reflection/transmission at the nucleus surface. Collisions can be elastic or inelastic. Most of the time experimental data (cross sections) are used to define interaction probabilities first and second what are the output products and their characteristics (types, energies, momenta), each selection being done randomly. When necessary, Pauli blocking is taken into account. Cross-section parameterizations, number and type of collisions and of resonances taken into account, and the way to apply Pauli blocking different for different models.

When the particle reaches the surface; it can be emitted or reflected. To be emitted the particle must be energetic enough, i.e. be able to overcome nuclear and Coulomb potential. However, the basics of INC are clearly the treatment of the transport of nucleons with their two-body interactions, i.e. without clusterization. So, up to now,, the only way to produce composite particles during the cascade is to add a coalescence model. Before leaving the nucleus, a nucleon can drag and aggregate one or more nucleons, close enough to it in space and momentum. This procedure extends the INC applicability to a satisfactory level.

Finally, different criteria are used to stop the cascade and to start the de-excitation phase of the remnant nucleus. We can mention three of them: cutoff energy, stopping time and deviation from an optical absorptive potential.

Different scenarios for intranuclear cascade are possible depending on the energy and the impact parameter of the incident particle going from ejection of a single nucleon, taking with it all of the incident energy, to the capture of the projectile leaving the nucleus in a state of strong excitation. Once excited, the nucleus enters a second and slower phase, the de-excitation. Here again, different scenarios compete according to the mass, excitation energy and angular momentum of the remnant nucleus. The first and rapid phase is of about $10^{-22}$ s and the second in the order of $10^{-18}$ s. In addition to these two phases sometimes included is a third one named pre-equilibrium. This step is actually an intermediate step since it deals with the transition between cascade and de-excitation and more precisely how the cascade is stopped. The need of this additional phase is then strongly connected with the cascade modeling.

INC reproduces successfully a wide variety of experimental data of hadron and pion induced reactions, using a small number of adjustable parameters, most with clear physical meaning.

I. LIEGE INTRANUCLEAR-CASCADE MODEL INCL4.6

The original Liege INC model for nucleon-induced reactions is described in [3, 4]. The standard INCL4.2 model is described in detail in [5, 6] and in references cited therein. The INCL4 model is a time-like intranuclear cascade model. In the initial state, all nucleons are prepared in phase space. Target nucleons are given position and momentum at random in agreement with Saxon-Woods and Fermi sphere distributions, respectively. They are moving in a constant potential well describing the nuclear mean field. The incident particle (nucleon or pion) is given the appropriate energy and an impact parameter at random. In this version, incident light particles (up to alphas) are viewed as a collection of on-shell nucleons, with a Fermi motion inside their reference frame, and with a total energy equal to the nominal total incident energy. The collision mechanism is assumed to proceed from a succession of binary collisions (and decays) well separated in space and time. The fate of all particles is followed as time evolves. The particles travel along a straight-line trajectories until two of them reach their minimum distance of approach, in which case they can be scattered provided the value of this distance is small enough, or until they hit the border of the potential well, supposed to describe the nuclear target mean field. Initial positions of target nucleons are taken at random in the spherical nuclear target volume with a sharp surface, initial momenta are generated stochastically in a Fermi sphere. Inelastic collisions, pion production, and absorption are supposed to appear and disappear through the $NN \rightleftharpoons N\Delta$ and $\Delta \rightleftharpoons \pi N$ reactions. For $\pi N$ interaction, experimental cross sections are uses, including nonresonant scattering, but the latter is treated as proceeding through the formation of a $\Delta$ with a very short lifetime; inelastic $\pi N$ scattering is neglected for convenience. In the $NN \to N\Delta$ process, the $\Delta$ particle is given the mass $m_\Delta$, taken at random from the distribution

$$f(m_\Delta) = F_N \frac{q^3}{q^3+q_0^3} \frac{1}{1+4\left(\frac{m_\Delta-m_\Delta^0}{\Gamma_0}\right)^2}, (1)$$

$$q^2 = \frac{(m_\Delta^2-(m_N-m_\pi)^2)(m_\Delta^2-(m_N+m_\pi)^2)}{4m_\Delta^2}, \quad (2)$$

with $m_\Delta$ lying in the interval $[m_N + m_\pi, \sqrt{s} - m_N]$, $\sqrt{s}$ being the center of mass (c.m.) energy of the collision, and consistent with energy-momentum conservation. The quantity $F_N$ in Eq. (1) is the normalization constant. The parameters are $q_0 = 0.18$ GeV, $m_\Delta^0 = 1.215$ GeV, and $\Gamma_0 = 0.13$ GeV. The introduction of the q-dependent factor is required by the fit of $NN \rightleftharpoons N\Delta$ data and can be justified as follows: a $\Delta$ resonance is a correlated pion-nucleon system and the phase space of the latter system is considerably reduced when its c.m. energy is low. The average intrinsic lifetime $\tau_\Delta$ was taken as follows:

$$\frac{1}{\tau_\Delta} = \frac{q^3}{q^3+q_0^3}\frac{1}{\tau_0}, \quad (3)$$

where proper time $\tau_0 = \frac{\hbar}{\Gamma_0}$. This is justified as follows: if the $\Delta$ resonance is going to decay into a $\pi N$ pair with low energy (which is the case for small $m_\Delta$ in our classical picture), the decay width is considerably diminished due the reduction of phase space. The stopping time of the cascade is determined self-consistently by the model itself. It can simply be parameterized (in fm/c) by

$$t_{stop} = 29.8 A_T^{0.16}, \quad (4)$$

for incident nucleons($Z_T$ and $A_T$ are the charge and mass numbers of the target, respectively). At the beginning of the cascade process, the incident nucleon or pion is located with its own impact parameter on the surface of the working sphere, which is centered on the target with the radius

$$R_{max} = R_0 + 8a, \quad (5)$$

where $R_0$ and a are respectively the radius and the diffuseness of the target nucleus density. Particles are moving along straight-line trajectories between collisions inside the working sphere. They are divided into participants and spectators in the usual sense. When participants leave the working sphere, they are considered as ejectiles and do not interact anymore. The potential radius for particles with energy larger than the Fermi energy is also taken to be equal to $R_{max}$. Pions do not experience any potential. The depth of the potential well felt by the nucleons is dependent on the energy of the nucleons and is not the same for protons and neutrons. The energy dependence is taken from the phenomenology of the real part of the optical-model potential.

$$\rho(r) = \begin{cases} \frac{\rho_0}{1+exp(\frac{(r-R_0)}{a})}, & for\ r < R_{max} \\ 0, & for\ r > R_{max} \end{cases}. \quad (6)$$

The values of $R_0$ and a are taken from electron scattering measurements and parameterized, for convenience, from $Al$ to $U$, as follows: $R_0 = (2.745\text{x}10^{-4}A_T + 1.063)(A_T)^{1/3}$ fm, $a = 0.510 + 1.63\text{x}10^{-4}A_T$ fm (in the numerical code, other values, as well as another shape for $\rho(r)$, can optionally be introduced). The quantity $\rho_0$ is such that the distribution is normalized to $A_T$, the target mass number. The momentum distribution is kept as a uniform Fermi sphere with Fermi momentum $p_F$. Nucleons with high momentum in the central part of the nucleus are expected to travel farther out than those with low momentum.

Therefore it is considered that a nucleon with the momentum $p$ is to reach the maximum radial distance $R(p)$. Because of these r-p correlations, a nucleon with the momentum between $p$ and $p + dp$ should be located, with a constant uniform probability, in a sphere of the radius $R(p)$. This radius can be deduced by assuming that the number of nucleons populating the layer of density profile $\rho(R(p))$ and $\rho(R(p + dp))$ is the same as the number of nucleons with the momentum between $p$ and $p + dp$:

$$A_T \frac{4\pi p^2 dp}{\frac{4\pi}{3} p_F^3} = \frac{4\pi}{3} R^3(p) dp. \quad (7)$$

The limiting conditions are naturally set to $R(0) = 0$ and $R(p_F) = R_{max}$, and the integration of Eq. (7) yields

$$\left(\frac{p}{p_F}\right)^{1/3} = -\frac{4\pi}{3A_T} \int_0^{R(p)} \frac{d\rho(r)}{dr} r^3 dr, \quad (8)$$

from which $R(p)$ can be deduced. The initial position and momentum of any target nucleon are generated as follows: $\vec{p}$ is taken at random in a sphere with the radius $p_F$, $R(p)$ is calculated using Eq. (8), and $\vec{r}$ is chosen at random in a sphere with the radius $R(p)$. This is equivalent to taking $\vec{r}$, $\vec{p}$ at random according to the joint probability distribution

$$\frac{dn}{d^3\vec{p}d^3\vec{r}} = f(\vec{r},\vec{p}) = A_T \frac{\theta(R(p)-r)\theta(p_F-p)}{\left(\frac{4\pi}{3}\right)^2 R^3(p) p_F^3}, \quad (9)$$

where $\theta(x)$ is the Heaviside function. In practice, the value of p can be generated from the uniform Fermi sphere distribution and the position is generated uniformly in a sphere with the radius $R(p)$. After integration over the relevant variables, the joint distribution in Eq. (9) corresponds to the spatial density $\rho(\vec{r})$ and to the sharp Fermi sphere momentum distribution:

$$\int f(\vec{r},\vec{p}) d^3p = \rho(\vec{r}), \quad (10)$$

$$\int f(\vec{r},\vec{p}) d^3r = A_T \frac{\theta(p_F-p)}{\frac{4\pi}{3} p_F^3}. \quad (11)$$

The procedure outlined above is at variance with the one used in many transport models, where nucleons are placed in a potential with a Saxon-Woods or similar shape. The dynamical Pauli blocking in INCL4.2 operates in phase space and is implemented as follows: if two nucleons $i$ and $j$ are going to suffer a collision at positions $\overrightarrow{r_{i(j)}}$ leading to the final state with momenta $\overrightarrow{p_{i(j)}}$, the phase space occupation probabilities $f_i$ are evaluated by counting nearby nucleons in a small phase-space volume,

$$f_i = \frac{1}{2} \frac{2\pi \hbar^3}{\frac{4\pi}{3} r_{PB}^3 p_{PB}^3} \times \Sigma_{k \neq j} \, \theta(r_{PB} - |\vec{r_k} - \vec{r_j}|) \times \theta(p_{PB} - |\vec{p_k} - \vec{p_j}|), \quad (12)$$

where the sum is limited to particles k with the same isospin component as particle i (or j). The factor ½ is introduced because spin components are ignored. The parameters $r_{PB}$ and $p_{PB}$ were taken to have the following values: $r_{PB}$ = 3.18 fm and $p_{PB}$ = 200 MeV/c. The collision between i and j is allowed or forbidden following the comparison of a random number with the product (1 $-f_i$)(1 $- f_j$ ). Pauli blocking is not applied to Δ particles because their density is always very small. On the other hand, it is enforced for nucleons resulting from Δ decays. At the end of the cascade, surviving Δ resonances from inelastic collisions are forced to decay and the conservation of baryon number, charge, energy, momentum and angular momentum is verified,

$A_P + A_T = A_{ej} + A_{rem}$, (13)

$Z_P + Z_T = Z_{ej} + Z_\pi + Z_{rem}$, (14)

$T_{lab} = K_{ej} + W_\pi + E_{rec} + E^* + S$, (15)

$\vec{p_{lab}} = \vec{p_{ej}} + \vec{p_\pi} + \vec{p_{rem}}$, (16)

$\vec{l} = \vec{l_{ej}} + \vec{l_\pi} + \vec{l_{rem}} + \vec{l^*}$, (17)

for baryon number, charge, energy, momentum, and angular momentum, respectively. The projectile P colliding with the target T and generating (baryonic) ejectiles, pions, and a remnant (which is the remaining part of the target up to the end of the cascade stage) are considered. In Eq. (15), $K_{ej}$ is the kinetic energy of the ejectiles, $W_p$ is the total energy of the pions, $E_{rec}$ is the recoil energy of the remnant, $E^*$ is the remnant excitation energy, and S is the separation energy, i.e., the minimum energy needed to remove all ejectiles and pions from the target nucleus ground state. In the other equations, the indices have the similar meaning. In Eq. (17), $\vec{l}$ is the angular momentum of the incident particle, $\vec{l_{rem}}$ is the angular momentum corresponding to the c.m. motion of the remnant, and $\vec{l^*}$ is the intrinsic angular momentum of the remnant. The INCL4.2 version was tested successfully, in the 200 MeV - 2 GeV range, against a large data base, but some phenomenological aspects of nuclear physics were neglected. The model cannot describe production of clusters in the cascade, i.e. with a kinetic energy definitely larger than the typical evaporation energies, as it can be seen experimentally. Concerning the predictive power of the model, several deficiencies can be noted. Pion production is generally overestimated. Quasi-elastic peaks in (p, n) reactions are generally too narrow and sometimes underestimated. Finally, the reaction cross sections are severely underestimated below ~100 MeV. Residue production cross sections are sometimes unsatisfactorily reproduced, especially for residues close to the target.

In 2009, an upgraded version (INCL4.5) was released [6] where the following model improvements were implemented.

An average isospin-dependent potential well, of the Lane type, is introduced for pions, as well as reflection and/or transmission at the border of this potential. The depth of the potential was taken as far as possible from the phenomenology of the real part of the pion-nucleus optical potential (dispersive effects due to the strong imaginary part have to be removed). This depth amounts to 22 MeV for $\pi+$ and 38 MeV for $\pi-$ on the $Pb$ target. The radius of the potential is taken as $R_0+4a$, in rough accordance with phenomenology. This modification reduces the pion production cross section, thus compensating the overestimation by INCL4.2.

An improvement of the INCL4.2 model is that outgoing nucleon crossing the nuclear periphery is supposed to be able to carry along other nucleons to form a cluster, provided the involved nucleons are lying sufficiently near each other in phase space. To limitations in computing time, clusters up to $A_{cl}^{max} = 8$ are considered. The emitted cluster should have sufficient energy to escape, i.e. $T_{cl} = \sum(T_i - V_i) - B_{cl} > 0$, where the $T_i$ are the kinetic energies of the nucleons and $V_i$ are the depths of their potential wells and the cluster has also to succeed the test for penetration through the Coulomb barrier. At the end of the cascade process, short-lived clusters with a lifetime of less than 1 ms (e.g. $^5Li$) are forced to decay, isotropically in their c.m. frame. Clusters with a lifetime larger than 1 ms are considered detectable, prior to decay.

In this way, the following features are introduced. First, the deflection of charged particles in the Coulomb field was taken into account, both for incident and outgoing particle at the nuclear periphery. Second, light charged particles can be emitted by the cascade owing to a dynamical coalescence model: nucleons leaving the target may carry other nucleons provided they are sufficiently close by in phase space. Third, the behavior of the model at low incident energy is improved, mainly by a better account of soft collisions, especially in the first instances of the reaction process.

Finally, the last version of the model, INCL4.6, was published in 2013. A detailed account can be found in [7]. The main new development involves the treatment of cluster-induced reactions.

Treatment of cluster-induced reactions is as follows. In INCL4.2, an incident cluster (up to an alpha particle) is considered as a collection of independent nucleons with internal Fermi motion superimposed to the motion of the incident cluster as a whole (see [5]), adjusted in such a way that the sum of the total energies of the constituting nucleons is equal to the nominal energy of the physical cluster. In other words, the cluster is replaced by independent on-shell nucleons with the correct nominal total energy, but with an incorrect (smaller than nominal) total momentum. This approximation is justified at high energy, but it is not really appropriate for reactions at low incident energy.

Initialization of the incident cluster is as follows. Nucleon momenta $\vec{p_i'}$ and positions $\vec{r_i'}$ i inside the cluster are generated as before [5] (note, however, that a special method is applied to ensure $\sum \vec{p_i'} = 0$ and $\sum \vec{r_i'} = 0$). At the beginning of the event, the cluster center of mass is positioned on the classical Coulomb trajectory in such a way that one of the nucleons is touching a sphere of radius $R_{Coul}$. The latter represents the Coulomb barrier. The value of $R_{Coul}$ is taken from the phenomenology of the Coulomb barrier heights and has been tabulated as function of the target mass for *p, d, t, ³He* and *⁴He* projectiles.

Collisions are, of course, governed by Pauli blocking, treated in a different way in the first and in the subsequent collisions. The nucleons involved in the first collision are subject to a strict blocking: after the collision, both of them should lie outside the Fermi sphere. In subsequent collisions, the blocking is applied stochastically, with a probability given by the product of final state blocking factors. A careful definition of the latter allows one to account for surface effects and for the depletion of the Fermi sphere during the evolution of the cascade.

An important novelty of recent versions of the code is the introduction of a coalescence model based on phase space, which permits the emission of light clusters, with mass $A \leq 8$, during the cascade stage, in keeping with experimental evidence.

The INCL4.6 version uses modified value for $R_{max}$

$R_{max} = R_0 + 8a + r_{int} = R_1 + 8a$, (18)

and the model separation energy $S_i$ is replaced by the physical separation energy $S_i^{phys}$, taken from mass tables, for the emission from the actual nucleus. The modification of $R_{max}$ increased the maximum time of the cascade, which now corresponds to the time of passage of the incident particle through the "working sphere" along a diameter, when this time exceeds the usual stopping time, given by Eq. (4).

An important characteristic of the model is the self-consistent determination of the stopping time of the cascade, which can be simply parameterized as $t_{stop} = 29.8 A_T^{0.16}$ fm/s, where $A_T$ is the mass of the target nucleus. At $t = t_{stop}$ many physical quantities, such as the excitation energy of the target nucleus and the average kinetic energy of the ejectiles, switch from a fast time evolution, dominated by intranuclear cascade, to a much slower evolution, which is taken as a signature of equilibration. Thanks to this choice of the stopping time, it is not necessary to introduce a pre-equilibrium model describing the intermediate stage between the fast cascade and the evaporation-fission decay.

Intranuclear-cascade models in general (and INCL in particular) only describe the fast, dynamical stage of a spallation reaction, leading to the formation of excited nuclei which subsequently de-excite by emitting particles and/or fissioning. It is therefore necessary to follow the de-excitation of this cascade remnant if one requires a complete description of the nuclear reaction. Since the time scale for de-excitation is much longer than for cascade, a different physical description is usually employed. This may include an optional pre-equilibrium stage, which then handles the thermalization of the remnant; if pre-equilibrium is used, the intranuclear-cascade stage is stopped earlier. Either way, thermalization is attained and subsequent de-excitation of the remnant is described by statistical de-excitation models.

Such a pre-equilibrium model is sometimes used between the cascade and the de-excitation phases. Several versions exist, but almost all are based on the exciton model developed by Griffin [8]. According to the use or rejection of this intermediate phase, the duration of the cascade is obviously different or, maybe more correctly, mass, charge and excitation of the remnant nucleus are larger, if this phase is called. While some intranuclear cascade models need such pre-equilibrium models to improve their capability, this is not the case of some others.

Boudard, co-developer of INCL, initiated the translation of the Fortran77 version of INCL in C++. The work, started by P. Kaitaniemi [11], was continued and finalized by D. Mancusi [12]. This provided the opportunity to re-consider the INCL code and made its maintenance easier. The main transport code implemented the INCL4.6 [7] version as the default intranuclear cascade model.

For this purpose, the Liege intranuclear cascade model (INCL) [12] is used; this model has been recently extended towards high energies ($\approx$ 15 GeV) including multipion production[13, 14], strange particles, such as kaons and hyperons [15, 16], and the production of $\eta$ and $\omega$ mesons [17]. This new version of the INCL allows us to predict the formation of hyperremnants and their characterization in atomic ,mass, and strangeness numbers together with their excitation energies and angular momenta. These improvements in INCL also require de-excitation models considering the emission of hyperons, in particular, the evaporation of particles. Currently, there are a few numbers of de-excitation models that treat the evaporation of hyperons and the formation of hypernuclei, such as the evaporation model ABLA07 developed at GSI by Kelic and collaborators [18] and recently extended to hypernuclei by us including the evaporation of particles on the basis of Weisskopfs approach according to [19].

II. BINARY CASCADE

Binary Cascade is a hybrid between a classical intranuclear cascade and a QMD[20] model, for the simulation of inelastic scattering of pions, protons and neutrons, and light ions of intermediate energies off nuclei[21]. The nucleus is modeled by individual nucleons bound in the nuclear potential. Binary collisions of projectiles or projectile constituents and secondaries with single nucleons, resonance production, and decay are simulated according to measured, parameterized or calculated cross sections. The Pauli exclusion principle, i.e. blocking of interactions due to Fermi statistics, reduces the free cross section to an effective intra-nuclear cross section. Secondary particles are allowed to further interact with remaining nucleons. The Binary Cascade models interactions of nucleons, pions, and light ions with nuclei for incident particle energies in the energy range starting from few MeV up to few GeV.

The Binary Cascade introduces a new approach to cascade calculations. It is based on a detailed 3-dimensional model of the nucleus, and exclusively based on binary scattering between reaction participants and nucleons within this nuclear model. This feature makes it a hybrid between a classical cascade code, and a quantum molecular dynamics model (QMD) [20]. In Binary Cascade, like in QMD, each participating nucleon is seen as a Gaussian wave package,

$$\phi(x; q_i; p_i; t) = 2/(L\pi)^{3/4} \exp\left(-2/L(x - q(t))^2 + ip_i(t)x\right), \quad (19)$$

propagating in time and space, undergoing collisions with nucleons in the nuclear medium in the process. Here, $x$ and $t$ are space and time coordinates, and $q_i$ and $p_i$ describe the nucleon position in the configuration and momentum space. The total wave function is assumed to be the direct product of the wave functions of the participating nucleons and hadrons. Participating means that they are either primary particles, or were generated or scattered in the process of the cascade.

Binary Cascade is an intra-nuclear cascade propagating primary nucleons and all secondary particles within a nucleus. Interactions take place between a primary or secondary particle and an individual nucleon of the nucleus. The nucleus is modeled by explicitly positioning nucleons in space, and assigning momenta to these nucleons. This is done in a way consistent with the nuclear density distributions, Pauli's exclusion principle, and the total nu- clear mass.

Propagating particles in the nuclear field is done by numerically solving the equations of motion, using time-independent fields derived from optical potentials. The cascade begins with a projectile and the nuclear description, and terminates when the average energy of all participants within the nuclear boundaries are below a given threshold. The remaining pre-fragment will be treated by pre-equilibrium decay and de-excitation models.

For the primary particle an impact parameter is chosen randomly on a disk outside the nucleus, perpendicular to a vector passing through the center of the nucleus. The initial direction of the primary is perpendicular to this disk. Using straight-line transport, the distance of closest approach $d_i^{min}$ to each nucleon $i$ in the target nucleus, and the corresponding time-of-flight $t_i^d$ is calculated. The interaction cross-section $\sigma_i$ with target nucleons is calculated based on the momenta of the nucleons in the nucleus, and the projectile momentum. The target nucleons for which the distance of the closest approach $d_i^{min}$ is smaller than $\sqrt{\frac{\sigma_i}{\pi}}$ are candidate collision partners for the primary. All candidate collisions are ordered by increasing $t_i^d$. If no collision is found, a new impact parameter is chosen. This way transparency effects at the nuclear boundaries are taken into account. The primary particle is then transported in the nuclear field by the time step given by the time to closest approach for the earliest collision candidate. Outside the nucleus, particles travel along straight-line trajectories. Particles entering the nucleus have their energy corrected for Coulomb effects. Inside the nucleus particles are propagated in the nuclear field. The equation of motion in the field is solved for a given time step using a Runge-Kutta integration method. At the end of each step, the interaction of the collision partners is simulated using the scattering term described below, resulting in a set of candidate particles for further transport. The secondaries from a binary collision are accepted subject to Pauli's exclusion principle. If the momentum of any of the particles is below the Fermi momentum, the interaction is suppressed, and the original primary continues to the time of its next collision. In case an interaction is Pauli allowed, the tracking of the primary ends, and the secondaries are treated like the primary. All their possible binary collisions with the residual nucleus are calculated, with the addition of decay in case of strong resonances. For resonance decay, the collision time is the time to the decay of the particle, sampled from the resonance's lifetime. Herein the stochastic masses and decay widths are taken into account. All secondaries are tracked until they react, decay or leave the nucleus, or until the cascade stops due to the cut-off condition described above.

A 3-dimensional model of the nucleus is constructed from $A$ nucleons and $Z$ protons with coordinates $r_i$ and momenta $p_i$, with i = 1, 2, . . . ,$A$. Nucleon radii $r_i$ are selected randomly in the nucleus rest frame according to the nuclear density $\rho(r_i)$. For nuclei with $A > 16$, the Woods-Saxon form of the nucleon density [22] is used,

$$\rho(r_i) = \frac{\rho_0}{1+\exp\left[(r_i-R)/a\right]}, \quad (20)$$

where $\rho_0$ is approximated as

$$\rho_0 = \frac{3}{4\pi R^3}\left(1 + \frac{a^2}{\pi^2 R^2}\right)^{-1}, \quad (21)$$

Here $a = 0.545$ fm, and $R = r_0 A^{\frac{1}{3}}$ fm with the correction $r_0 = 1.16(1 - 1.16A^{-\frac{2}{3}})$ fm. For light nuclei, the harmonic-oscillator shell model for the nuclear density [23] is used,

$$\rho(r_i) = (\pi R^2)^{-3/2} \exp\left(\frac{-r_i^2}{R^2}\right), \quad (22)$$

where $R^2 = \frac{2}{3}\langle r^2 \rangle = 0.8133 A^{2/3} \, fm^2$. To take into account the repulsive core of the nucleon-nucleon potential, a minimum inter-nucleon distance of 0.8 fm is taken. The nucleus is assumed to be spherical and isotropic, i.e. each nucleon is placed using a random direction and the previously determined radius $r_i$.

The momenta $p_i$ of the nucleons are chosen randomly between 0 and the Fermi momentum $p_F^{max}(r_i)$. The Fermi momentum, in the local Thomas-Fermi approximation as a function of the nuclear density $\rho$, is

$$p_F^{max}(r_i) = \hbar c(3\pi^2 \rho(r))^{1/3}. \quad (23)$$

The total vector sum of the nucleon momenta has to be zero, i.e. the nucleus must be constructed at rest. To achieve this, one nucleon is chosen to compensate the vector sum of the remaining nucleon momenta $p_{rest} = -\sum_{i=1}^{i=A-1} p_i$. If this sum is larger than the maximum allowable momentum $p_F^{max}(r_i)$, the direction of the momenta of the nucleons with the largest contribution to the net nucleus momentum is iteratively flipped, until the residual sum is an allowed momentum value for a nucleon.

The effect of collective nuclear interaction upon participants is approximated by a time-invariant scalar optical potential, based on the properties of the target nucleus. For protons and neutrons the potential used is determined by the local Fermi momentum $p_F(r)$ as

$$V(r) = \frac{p_F^2}{2m}, (24)$$

where *m* is the mass of the neutron or the mass of the proton, respectively. For pions the potential used is a simple approximation given by the lowest-order optical potential as derived in [24]:

$$V(r) = -\frac{-2\pi(\hbar c)^2 A}{\overline{m}_\pi}\left(1 + \frac{m_\pi}{M}\right) b_0 \rho(r). (25)$$

Here, A is the nuclear mass number, and $m_\pi$ and M are the pion and nucleon masses, respectively; $\overline{m}_\pi$ is the reduced pion mass, $\overline{m}_\pi = \frac{m_\pi + m_N}{m_\pi M}$, where $m_N$ is the mass of the nucleus; *ρ(r)* is the nucleon density distribution. The parameter $b_0$ is the effective s-wave scattering length. The value used was obtained from the analysis to pion atomic data and resulted in $b_0$ to be about -0.042 fm. It is assumed that the nucleus is in its ground state and all states below Fermi energy are occupied. Thus, collisions and decays for which any secondary nucleon has a momentum pi below the local Fermi momentum, i.e.

$$p_i < p_F^{max}(r_i), (26)$$

are suppressed.

The basis of the description of the reactive part of the scattering amplitude are two-particle binary collisions, also with associated or direct resonance production, and decay. Based on the cross section described later, collisions will occur when the transverse distance $d_t$ of any participant target pair becomes smaller than the black-disk radius corresponding to the total cross-section $\sigma_t$

$$\frac{\sigma_t}{p_i} > d_t^2. (27)$$

Experimental data and parameterizations thereof are used in the calculation of the total, inelastic and elastic cross-section wherever available.

For the case of proton-proton (*pp*) and proton-neutron (*pn*) collisions, as well as $\pi^+$- and $\pi^-$-nucleon collisions, experimental data and parameterizations are readily available as collected by the Particle Data Group (PDG) [25] for both elastic and inelastic collisions. The tabulation based on a sub-set of these data for $\sqrt{s}$ below 3 GeV, and the PDG parameterization at higher energies, are applied.

It also defines an upper limit of applicability of the model. Below 10 GeV kinetic energy, the resonance contributions considered are wholly sufficient to describe the total cross-section.

Most of the cross-sections of individual channels involving meson-nucleon scattering can be modeled as resonance excitation in the s-channel.

The initial states included in the model at present include all pion-nucleon scattering channels. The product resonances taken into account are the Delta-resonances with masses of 1232, 1600, 1620, 1700, 1900, 1905, 1910, 1920, 1930, and 1950 MeV, and the excited nucleons with masses of 1440, 1520, 1535, 1650, 1675, 1680, 1700, 1710, 1720, 1900, 1990, 2090, 2190, 2220, and 2250 MeV.

In the resonance production in the t-channel, single and double resonance excitations in nucleon-nucleon collisions are taken into account.

The resonance production cross-sections are as much as possible based on parameterizations of experimental data for proton-proton scattering. The formula used for parameterizing the cross-sections is motivated from the form of the exclusive production cross- section of the $\Delta_{1232}$ in proton-proton collisions:

$$\sigma_{AB} = 2\alpha_{AB}\beta_{AB} \frac{\sqrt{s}-\sqrt{s_0}}{(\sqrt{s}-\sqrt{s_0})^2+\beta_{AB}^2} x \left(\frac{\sqrt{s_0}+\beta_{AB}}{\sqrt{s}}\right)^{\gamma}_{AB}. \quad (28)$$

For all other channels, the parameterizations were derived from these by adjusting the threshold behavior accordingly. Cross-sections for the reminder of the channels are derived from those described above, by applying detailed balance. Isospin invariance is assumed. The formalism used to apply detailed balance is

$$\sigma(cd \to ab) = \sum_{J,M} \frac{\langle j_c m_c j_d m_d | JM \rangle^2}{\langle j_a m_a j_b m_b | JM \rangle^2} \frac{(2S_a+1)(2S_b+1)}{(2S_c+1)(2S_d+1)} \frac{\langle p_{AB}^2 \rangle}{\langle p_{CD}^2 \rangle} \sigma(ab \to cd). \quad (29)$$

Angular distributions for elastic scattering of nucleons are taken as closely as possible from experimental data, i.e. from the result of a phase shift analysis. They are derived from differential cross-sections obtained from the SAID database, R. Arndt et al. [26].

Angular distributions for final states other than nucleon-nucleon elastic scattering are calculated analytically, derived from the collision term of the in-medium relativistic Boltzmann-Uehling-Uhlenbeck equation [27] via scaling of the center-of-mass energy. The modular structure of GEANT4 allows the generation of single events with a known incident particle energy and any explicitly defined hadronic final-state generator. The kinematics of secondaries produced in the interaction are then analyzed and the resulting angular, momentum, energy, and baryon number spectra are stored in histograms. The energy-momentum balance can be controlled as well. The histograms are compared to published measurements of the differential and double differential $d\sigma/dE$ cross sections, $d\sigma/dE$, $d\sigma/d\Omega$, $d^2\sigma/dEd\Omega$, and the invariant cross-sections, $Ed^3\sigma/d^3p$.

The range of Binary Cascade model applicability in nucleon nuclear reactions stretches from < 100 MeV to about 10 GeV, allowing for a consistent calculation of the secondary hadron spectra in the low and intermediate energy domains.

III. BERTINI CASCADE

The INC model developed by Bertini [28–31] solves on the average the Boltzmann equation of this particle interaction problem. The Bertini nuclear model consists of a three-region approximation to the continuously changing density distribution of nuclear matter within nuclei. Relativistic kinematics is applied throughout the cascade and the cascade is stopped when all the particles which can escape the nucleus, have done so. The Pauli exclusion principle is taken into account and conformity with the energy conservation law is checked. Path lengths of nucleons in the nucleus are sampled according to the local density and free nucleon-nucleon cross-sections. Angles after collisions are sampled from experimental differential cross-sections. Intermediate energy nuclear reactions up to 10 GeV energy are treated for proton, neutron, pions, photon and nuclear isotopes.

The necessary condition of validity of the INC model is $\lambda_B/v \ll \tau_c \ll \Delta t$, where $\lambda_B$ is the de Broglie wavelength of the nucleons, $v$ is the average relative nucleon-nucleon velocity and $\Delta t$ is the time interval between collisions. The physical foundation becomes approximate at energies less than about 200 MeV, and there needs to be supplemented with a pre-equilibrium model. Also, at energies higher than 5-10 GeV the INC picture breaks down. The basic steps of the INC model are summarized below.

The nucleons are assumed to have a Fermi gas momentum distribution. The Fermi energy is calculated in a local density approximation i.e. it is made radius dependent with the Fermi momentum $p_F(r) = \left(\frac{3\pi^2 \rho(r)}{2}\right)^{1/3}$. The initialization phase fixes the nucleus radius and momentum according to the Fermi gas model.

If the target is Hydrogen ($A = 1$), a direct particle-particle collision is performed, and no nuclear modeling is used.

If $1 < A < 4$, a nuclei model consisting of one layer with a radius of 8.0 fm is created.

For $4 < A < 11$, a nuclei model is composed of three concentric spheres i = {1, 2, 3} with the radii

$$r_i(\alpha_i) = \sqrt{C_1^2 \left(1 - \frac{1}{A}\right) + 6.4\sqrt{-\log(\alpha_i)}},$$

where $\alpha_i$ = {0.01, 0.3, 0.7} and $C_1 = 3.3836 A^{1/3}$.

If A > 11, nuclei are modeled with three concentric spheres as well. The sphere radii are then defined as:

$$r_i(\alpha_i) = C_2 \log\left(\frac{1+e^{-\frac{C_1}{C_2}}}{\alpha_i} - 1\right) + C_1,$$

where $C_2 = 1.7234$.

The potential energy for nucleon N is

$$V_N = \frac{p_F^2}{2m_N} + BE_N(A,Z)$$

where $p_F$ is the Fermi momentum and BE the binding energy.

The momentum distribution in each region follows the Fermi distribution with zero temperature.

$$f(p) = cp^2, \qquad (30)$$

where

$$\int_0^{p_F} f(p)dp = n_p \text{ or } n_n. \qquad (31)$$

Here $n_p$ and $n_n$ are the numbers of protons and neutrons in the region and $p_F$ is momentum corresponding the Fermi energy

$$E_F = \frac{p_F^2}{2m_N} = \frac{\hbar^2}{2m_N}\left(\frac{3\pi^2}{v}\right)^{\frac{2}{3}}, \qquad (32)$$

which depends on the density $n/v$ of particles, and which is different for each particle and each region. The path lengths of nucleons in the nucleus are sampled according to the local density and free nucleon-nucleon cross-sections. The angles after collisions are sampled from experimental differential cross-sections.

Thus, the free particle-particle cross-sections and region-dependent nucleon densities are used to select the path length for the projectile particle. The tabulated total reaction cross-sections are calculated by Letaw's formulation [32–34]. For nucleon-nucleon cross-sections, parameterizations based on the experimental energy and isospin dependent data are used.

For pions the INC cross-sections are provided to treat elastic collisions, and inelastic channels: $\pi^- n \to \pi^0 n$, $\pi^0 p \to \pi^+ n$ and $\pi^0 n \to \pi^- p$. Multiple particle production is also implemented.

The S-wave pion absorption channels $\pi^+ nn \to pn$, $\pi^+ pn \to pp$, $\pi^0 nn \to X$, $\pi^0 pn \to pn$, $\pi^0 pp \to pp$, $\pi^- nn \to X$, $\pi^- pn \to nn$, and $\pi^- pp \to pn$ are implemented.

The Pauli exclusion principle forbids interactions where the products would be in occupied states. Following the assumption of a completely degenerate Fermi gas, the levels are filled from the lowest level. The minimum energy allowed for a collision product corresponds to the lowest unfilled level of system, which is the Fermi energy in the region. So, in practice, the Pauli exclusion principle is taken into account by accepting only secondary nucleons which have $E_N > E_F$.

After INC, the residual excitation energy of the resulting nucleus is used as input for a non-equilibrium model. The Geant4 cascade model implements the exciton model proposed by Griffin [8]. In this model nuclear states are characterized by the number of exited particles and holes (the exitons). INC collisions give rise to a sequence of states characterized by increasing exciton number, eventually leading to an equilibrated nucleus. For practical implementation of the exciton model we use level density parameters from [35] and the matrix elements from [36]. In the exciton model the possible selection rules for particle-hole configurations in the course of the cascade are: $\Delta p = 0,\pm 1$; $\Delta h = 0,\pm 1$; $\Delta n = 0,\pm 2$, where p is the number of particles, h is number of holes, and n = p + h is the number of exitons. The cascade pre-equilibrium model uses target excitation data, and exciton configurations for neutrons and protons to produce the non-equilibrium evaporation. The angular distribution is isotropic in the frame of rest of the exciton system. The parameterizations of the level density used are tabulated both with their *A* and *Z* dependence and including high temperature behavior. The smooth liquid high energy formula is used for the nuclear binding energy.

Fermi break-up is allowed only in some extreme cases, i.e. for light nuclei ($A < 12$ and $3(A - Z) < Z < 6$)

and if $E_{excitation} > 3E_{binding}$. A simple explosion model decays the nucleus into neutrons and protons and decreases exotic evaporation processes. The fission model is a phenomenological model using potential minimization. The binding energy parametrization is used and some features of the fission statistical model are incorporated as in [37].

The statistical theory for particle emission from exited nuclei remaining after INC was originally developed by Weisskopf [38]. This model assumes complete energy equilibration before particle emission, and re-equilibration of excitation energies between successive evaporation emissions. As a result, the angular distribution of emitted particles is isotropic.

The emission of particles is computed until the excitation energy falls below the cutoff value. If a light nucleus is highly exited, the Fermi break-up model is executed. In addition, fission is performed when the fission channel is open. The main chain of evaporation is followed until $E_{excitation}$ falls below $E_{cutoff} = 0.1$ MeV. The evaporation model ends with a $\gamma$ emission chain, which is followed until $E_{excitation} < E^{\gamma}{}_{cutoff} = 10^{-15}$ MeV.

Extensive benchmarking of the INC physics provided by Bertini cascade sub-models, exitons, pre-equilibrium state, nucleus explosion, fission, and evaporation has been made. The Geant4 evaporation model for cascade implementation adapts the widely used computational method developed by Dostrowski [39, 40]. The model is validated up to 10 GeV incident energy and users from various fields have been using it successfully.

To validate Bertini isotope production physics performance, extensive simulations on proton-induced reactions in Pb and Au targets were performed with Geant4 [41]. The Bertini cascade model in Geant4 simulates the hadronic interactions of protons, neutrons and pions with surrounding materials.

IV. CEM AND LAQGSM MODELS

The Los Alamos National Laboratory (LANL) Monte-Carlo N-particle transport code MCNP6 [42] uses by default the latest version of the cascade-exciton model (CEM), CEM03.03 [43–45], as its event generator to simulate reactions induced by nucleons, pions, and photons with energies up to 4.5 GeV and the Los Alamos version of the quark-gluon string model (LAQGSM), LAQGSM03.03 [45–47], to simulate such reactions at higher energies, as well as reactions induced by other elementary particles and by nuclei with energies up to ~ 1 TeV/nucleon. Details, examples of results, and useful references to different versions of CEM and LAQGSM can be found in [45].

The cascade-exciton model (CEM) [44] of nuclear reactions is based on the standard Dubna intranuclear cascade model [48, 49] and the modified exciton model (MEM) [50, 51]. The CEM code calculates nuclear reactions induced only by nucleons, pions, and photons. A detailed description of the initial version of the CEM can be found in [44], therefore we outline here only its basic assumptions. The CEM assumes that reactions occur in three stages. The first stage is the INC, in which primary particles can be re-scattered and produce secondary particles several times prior to absorption by, or escape from, the nucleus. All the cascade calculations are carried out in the three-dimensional geometry. The nuclear matter density $\rho(r)$ is described by the Fermi distribution with two parameters taken from the analysis of electron-nucleus scattering, namely

$$\rho(r) = \rho_p(r) + \rho_n(r) = \rho_0\{1 + exp[(r-c)/a]\}, (33)$$

where $c = 1.07A^{1/3}$ fm, $A$ is the mass number of the target, and $a = 0.545$ fm. For simplicity, the target nucleus is divided by concentric spheres into seven zones in which the nuclear density is considered to be constant. The energy spectrum of the target nucleons is estimated in the perfect Fermi-gas approximation with the local Fermi energy $T_F(r) = \hbar^2[3\pi^2\rho(r)]^{2/3}/(2\,m_N)$, where $m_N$ is the nucleon mass. The influence of intranuclear nucleons on the incoming projectile is taken into account by adding to its laboratory kinetic energy the effective real potential $V$, as well as by considering the Pauli principle which forbids a number of intranuclear collisions and effectively increases the mean free path of cascade particles inside the target. For incident nucleons $V \equiv V_N(r) = T_F(r) + \epsilon$, where $T_F(r)$ is the corresponding Fermi energy and $\epsilon$ is the binding energy of the nucleons. For pions, CEM03.01 uses a square-well nuclear potential with the depth $V_\pi \simeq 25$ MeV, independently of the nucleus and pion energy, as was done in the initial Dubna INC [48, 49].

The Pauli exclusion principle at the cascade stage of the reaction is handled by assuming that nucleons of the target occupy all the energy levels up to the Fermi energy. Each simulated elastic or inelastic interaction of the projectile (or of a cascade particle) with a nucleon of the target is considered forbidden if the "secondary" nucleons have energies smaller than the Fermi energy. If they do, the trajectory of the particle is traced further from the forbidden point and a new interaction point, a new partner and a new interaction mode are simulated for the traced particle, etc., until the Pauli principle is satisfied or the particle leaves the nucleus.

If the residual nuclei after the INC have atomic numbers with $A \leq A_{Fermi} = 12$, CEM uses the Fermi break-up model to calculate their further disintegration instead of using the pre-equilibrium and evaporation models. Fermi break-up, which estimates the probabilities of various final states by calculating the approximate phase space available for each configuration, is much faster to calculate and gives results very similar to those from using the continuation of the more detailed models for lighter nuclei.

An important ingredient of the CEM is the criterion for transition from the intranuclear cascade to the pre-equilibrium model. The cascade model uses a different criterion to decide when a primary particle is considered to have left the cascade (cutoff energy $T_{cut}$ or cutoff time $t_{cut}$). In CEM the effective local optical absorptive potential $W_{opt.mod.}(r)$ is defined from the local interaction cross section of the particle, including Pauli blocking effects. This imaginary potential is compared to the one defined by the phenomenological global optical model $W_{opt.exp.}(r)$. The degree of similarity or difference of these imaginary potentials is characterized by the parameter

$$\mathcal{P} = |(W_{opt.mod.} - W_{opt.exp.})/W_{opt.exp.}|.$$

When $\mathcal{P}$ increases above an empirically chosen value, the particle leaves the cascade, and is then considered to be an exciton. CEM uses the fixed value $\mathcal{P} = 0.3$.

When the cascade stage of a reaction is completed, CEM uses the coalescence model to create high-energy $d$, $t$, $^3He$, and $^4He$ fragments by final-state interactions among emitted cascade nucleons outside of the target nucleus. The value of the momentum $p$ of each cascade nucleon is calculated relativistically from its kinetic energy $T$. It is assumed that all the cascade nucleons having differences in their momenta smaller than $p_c$ and with the correct isotopic content form an appropriate composite particle.

The coalescence model first checks all nucleons to form 2-nucleon pairs, if their momenta permit it. It then takes these 2-nucleon pairs and the single nucleons left and forms $^4He$, $^3He$, and/or tritium, if their momenta permit it. The extended coalescence model further takes these two-nucleon pairs, tritium, $^3$He, and $^4$He to see if they can coalesce to form heavier clusters: $^6$He, $^6$Li, $^7$Li or $^7$Be. All coalesced nucleons are removed from the distributions of nucleons so that atomic and mass numbers are conserved.

The results show significant improvement in the production of heavy clusters in the whole energy range. However, too many alpha particles were lost (coalesced into heavy clusters); so $p_c(^4He)$ was increased to compensate it. The new values for $p_c$ for the extended coalescence model are:

$p_c(d) = 90\ MeV/c$;

$p_c(t) = p_c(^3He) = 108\ MeV/c$; (34)

$p_c(^4He) = 130\ MeV/c$;

$p_c(LF) = 175\ MeV/c$.

For 300 MeV < T < 1000 MeV they are:

$p_c(d) = 150\ MeV/c$;

$p_c(t) = p_c(^3He) = 175\ MeV/c$; (35)

$p_c(^4He) = 205\ MeV/c$;

$p_c(LF) = 250\ MeV/c$.

The emission of the cascade particles determines the particle-hole configuration, Z, A, and the excitation energy that is the starting point for the pre-equilibrium stage of the reaction. The subsequent relaxation of the nuclear excitation is treated in terms of an improved Modified Exciton Model (MEM) [50, 51] of pre-equilibrium decay, followed by the equilibrium evaporation/fission stage described using a modification of the generalized evaporation model (GEM) code GEM2 by Furihata [52]. The transition from the pre-equilibrium stage of a reaction to the third (evaporation) stage occurs when the probability of nuclear transitions changing the number of excitons n with Δn = +2 becomes equal to the probability of transitions in the opposite direction, with Δn = −2, i.e., when the exciton model predicts that equilibration has been established in the nucleus.

Generally, all three components can contribute to experimentally measured particle spectra and other distributions.

The Los Alamos version of the Quark-Gluon String Model (LAQGSM) [46, 47] is a further development of the Quark-Gluon String Model (QGSM) by Amelin, Gudima, and Toneev (see [55] and references therein) and is intended to describe both particle- and nucleus-induced reactions at energies up to about 1 TeV/nucleon. The basis of QGSM is the time-dependent version of the intranuclear-cascade model developed at Dubna, often referred in literature simply as the Dubna intranuclear Cascade Model (DCM) (see [53] and references therein). LAQGSM also describes nuclear reactions as three-stage processes: an INC, followed by pre-equilibrium emission of particles during the equilibration of the excited residual nuclei formed after the INC, followed by evaporation of particles from and/or fission of the compound nuclei.

The DCM models interactions of fast cascade particles ("participants") with nucleon spectators of both the target and projectile nuclei and includes as well interactions of two participants (cascade particles). It uses experimental particle+particle cross sections at energies below 4.5 GeV/nucleon, or those calculated by the quark-gluon string model (QGSM) at higher energies (see, e.g., [54] and references therein) to simulate angular and energy distributions of cascade particles, and also considers the Pauli exclusion principle.

When the cascade stage of a reaction is completed, QGSM uses the coalescence model described in [53] to "create" high-energy $d, t, {}^3He$, and ${}^4He$ by final-state interactions among emitted cascade nucleons outside of the colliding nuclei. After calculating the coalescence stage of a reaction, QGSM moves to the description of the last slow stages of the interaction, namely to pre-equilibrium decay and evaporation, with a possible competition of fission using the standard version of CEM [44]. If the residual nuclei have atomic numbers $A \leq 12$, QGSM uses the Fermi break-up model to calculate their further disintegration instead of using the pre-equilibrium and evaporation models. LAQGSM differs from QGSM by replacing the pre-equilibrium and evaporation parts of QGSM described according to the standard CEM [44] with the new physics from CEM2k [56, 57] and has a number of improvements and refinements in the cascade and Fermi break-up models. A detailed description of LAQGSM and further references can be found in [46, 47]. The coalescence model was extended to be able to produce light fragments up to ${}^7Be$ in CEM and up to ${}^{12}C$ in LAQGSM.

The pre-equilibrium interaction stage of nuclear reactions is considered by the current CEM and LAQGSM in the framework of the latest version of MEM [50, 51]. At the pre-equilibrium stage of a reaction, CEM03.03 and LAQGSM03.03 take into account all possible nuclear transitions changing the number of excitons $n$ with $\Delta n = +2, -2$, and 0, as well as all possible multiple subsequent emissions of $n, p, d, t, {}^3He$, and ${}^4He$. The corresponding system of master equations describing the behavior of a nucleus at the pre-equilibrium stage is solved by the Monte-Carlo method [44]. In [58], the modified exciton model MEM was extended to include the possibility of emitting heavy clusters, with $A > 4$, up to ${}^{28}Mg$ (66 types of particles and LF). For incident energies below about 200 MeV, Kalbach has developed a phenomenological systematics for pr-eequilibrium particle angular distributions by fitting available measured spectra of nucleons and complex particles [59]. As the Kalbach systematics are based on measured spectra, they describe very well the double-differential spectra of pre-equilibrium particles and generally provide a better agreement of calculated pre-equilibrium complex-particle spectra with experimental data.

The inverse cross sections used by these models at the pre-equilibrium stage (and at the evaporation/fission-stage) have a significant impact on the calculated particle width, and affect greatly the final results and the-accuracy of the MCNP6, MCNPX [60] and MARS15 [61-63]-transport codes, which use these models as their event-generators. This is why it is necessary to use as good as possible approximations for the inverse cross sections in the extended models.

The unmodified codes use the inverse cross sections $\sigma_{inv}$ from Dostrovsky's formulas [39, 40] for all emitted nucleons and complex particles ($d$, $t$, $^3He$, and $^4He$) is not very suitable for emission of fragments heavier than $^4He$. Better total-reaction-cross-section models that can be used as an estimate for inverse cross sections are available today, especially such as the NASA model [64], the approximations by Barashenkov and Polanski [65], and those by Kalbach [66]. A quite complete list of references on modern total-reaction-cross-section models, as well as on recent publications where these models are compared with each other and with available experimental data can be found in [67].

An extensive comparison of the systematics for total reaction (inverse) cross sections showed that the NASA approach is better, in general, than the other available models. This is why we implemented the NASA inverse cross sections into the MEM to be used at the pre-equilibrium stage of reactions.

The NASA approximation, as described by Eq. (36,) attempts to simulate several quantum-mechanical effects, such as the optical potential for neutrons (with the parameter $X_m$) and collective effects like Pauli blocking (through the quantity $\delta_T$).

$$\sigma_{NASA} = \pi r_0^2 \left(A_P^{1/3} + A_T^{1/3} + \delta_T\right)^2 \left(1 - R_c \frac{B_T}{T_{cm}}\right) X_m, \quad (36)$$

where $r_0$, $A_P$, $A_T$, $\delta_T$, $R_c$, $B_T$, $T_{cm}$, and $X_m$ are, respectively, the constant used to calculate the radii of nuclei, the mass number of the projectile nucleus, the mass number of the target nucleus, the energy-dependent parameter, the system-dependent Coulomb multiplier, the energy-dependent Coulomb barrier, the colliding system center-of-momentum energy, and the optical model multiplier used for neutron-induced reactions. The calculation of inverse cross sections at the pre-equilibrium stage of reactions was improved with a new hybrid NASA-Kalbach approach, instead of the old Dostrovsky model used previously. This extended version of the MEM is implemented into the upgraded CEM, labeled CEM03.03F, as well as into the new LAQGSM03.03F.

After the INC, LAQGSM uses the same pre-equilibrium, coalescence, Fermi break-up, and evaporation/fission models as described above for CEM.

The improved CEM, LAQGSM, was implemented as event generator into MCNP6 and allow one to describe particle- and nucleus-induced reactions and provide a good agreement with available experimental data. They have a good predictive power for various reactions and can be used as reliable tools in scientific and applied research.

Emission of energetic heavy clusters heavier than $^4He$ from nuclear reactions play a critical role in several applications, including electronics performance in space, human radiation dosages in space or other extreme radiation environments, proton and hadron therapy in medical physics, accelerator and shielding applications, and so on. The CEM and LAQGSM event generators in MCNP6 describe quite well the spectra of fragments with sizes up to $^4He$ in a broad range of target masses and incident energies (up to ~ 5 GeV for CEM and up to ~ 1 TeV/A for LAQGSM).

V. Numerical simulation of extended heavy targets irradiated by proton and deuteron beams. Comparison of models and codes

Thorough understanding of the mechanisms and approaches used in different simulation models is important for obtaining reliable numerical data on irradiation of big heavy targets by accelerated proton and ion beams. Such experiments performed at the JINR accelerator facilities with Quinta and BURAN targets contribute to the research aimed at advanced schemes of nuclear power production with accelerated particle beams. Of course, experimental studies of such scale and complexity should be preceded by comprehensive numerical study.

Below we give the results of simulation of extended heavy targets irradiated by proton and deuteron beams. The focus of comparison of different applied models is neutron production and absorption in the target material depending on the target dimensions.

We consider targets from uranium-238 with the following dimensions: a radius of 15 cm, a length of 40 cm; a radius of 30 cm, a length of 80 cm; and a radius of 60 cm, a length of 160 cm.

The following accelerated beams are considered: protons with an energy of 0.66 GeV, 1 GeV, 2 GeV, 4 GeV, and deuterons with an energy of 0.66 GeV/nucleon, 1 GeV/nucleon, 2 GeV/nucleon, 4 GeV/nucleon.

Among the most important parameters characterizing beam interaction with fissionable materials is neutron production rate per beam particle. Neutron flux leaving the target is also important, as it provides information on the energy accumulated in the target due to ion-target interaction.

Tables 1-12 below summarize the results of calculations via several models for the beam and target parameters given above.

Table 1. Neutron production/loss in interaction of p beam with U-238 target with a radius of 15 cm and a length of 40 cm obtained using the following models: a - SHIELD [68, 69]; b -

GEANT4 (BC-Bertini [28-31]), c - MCNP6 (CEM03 [43-45] and INCL [5-7]); d - MARS15 [61-63].

| $E_p$, GeV | Neutron production/part. | | | | | Neutron leakage/part. | | | | |
|---|---|---|---|---|---|---|---|---|---|---|
| | a | b | c | | d | a | b | c | | d |
| | | | CEM03 | INCL | | | | CEM03 | INCL | |
| 4 | 272 | 306 | 345 | 270 | 267 | 168 | 172 | 203 | 159 | 164 |
| 2 | 148 | 164 | 170 | 151 | 161 | 92 | 93 | 102 | 90 | 86 |
| 1 | 72 | 78 | 80 | 72 | 78 | 45 | 45 | 49 | 44 | 42 |
| 0.66 | 42 | 42 | 46 | 40 | 44 | 27 | 25 | 28 | 25 | 24 |

Table 2. Number of fissions and neutron captures per particle in interaction of p beam with U-238 target with a radius of 15 cm and a length of 40 cm obtained using the following models: a - SHIELD [68, 69];  b - GEANT4 (BC-Bertini [28-31]), c - MCNP6 (CEM03 [43-45] and INCL [5-7]); d - MARS15 [61-63].

| $E_p$, GeV | Number of fissions/part. | | | | | Number of captures/part. | | | | |
|---|---|---|---|---|---|---|---|---|---|---|
| | a | b | c | | d | a | b | c | | d |
| | | | CEM03 | INCL | | | | CEM03 | INCL | |
| 4 | 43.0 | 42.5 | 42 | 30 | 48 | 56 | 53.6 | 64 | 49 | 50,5 |
| 2 | 23.0 | 22.8 | 20 | 17 | 25 | 31 | 28.4 | 31 | 27 | 30 |
| 1 | 11.4 | 11.0 | 9 | 8 | 12 | 14.9 | 13.7 | 14 | 13 | 14.4 |
| 0.66 | 6.7 | 6.1 | 5 | 4 | 7 | 8.3 | 7.2 | 8 | 7 | 7.9 |

Table 3. Neutron production/loss in interaction of p beam with U-238 target with a radius of 30 cm and a length of 80 cm obtained using the following models: a - SHIELD [68, 69]; b - GEANT4 (BC-Bertini [28-31]), c - MCNP6 (CEM03 [43-45] and INCL [5-7]); d - MARS15 [61-63].

| $E_p$, GeV | Neutron production/part. | | | | | Neutron leakage/part. | | | | |
|---|---|---|---|---|---|---|---|---|---|---|
| | a | b | c | | d | a | b | c | | d |
| | | | CEM03 | INCL | | | | CEM03 | INCL | |
| 4 | 349 | 398 | 441 | 400 | 355 | 116 | 126 | 138 | 115 | 105 |
| 2 | 186 | 205 | 210 | 200 | 205 | 65 | 66 | 71 | 62 | 57 |
| 1 | 84 | 93 | 88 | 88 | 94 | 31 | 33 | 32 | 32 | 27 |
| 0.66 | 45 | 49 | 55 | 48 | 52 | 19 | 19 | 22 | 19 | 16 |

Table 4. Number of fissions and neutron captures per particle in interaction of p beam with U-238 target with a radius of 30 cm and a length of 80 cm obtained using the following models: a - SHIELD [68, 69];  b - GEANT4 (BC-Bertini [28-31]), c - MCNP6 (CEM03 [43-45] and INCL [5-7]); d - MARS15 [61-63].

| $E_p$, GeV | Number of fissions/part. | | | | | Number of captures/part. | | | | |
|---|---|---|---|---|---|---|---|---|---|---|
| | a | b | c | | d | a | b | c | | d |
| | | | CEM03 | INCL | | | | CEM03 | INCL | |
| 4 | 59 | 58.4 | 54 | 46 | 66 | 167 | 161.9 | 180 | 169 | 150.6 |
| 2 | 31 | 30.1 | 26 | 23 | 33 | 86 | 82.4 | 82 | 80 | 83.5 |
| 1 | 14 | 13.7 | 10 | 10 | 15 | 37 | 35.4 | 33 | 33 | 36.9 |
| 0.66 | 7 | 7.2 | 6 | 5 | 8 | 19 | 17.4 | 19 | 17 | 19.1 |

Table 5. Neutron production/loss in interaction of p beam with U-238 target with a radius of 60 cm and a length of 160 cm obtained using the following models: a - SHIELD [68, 69]; b - GEANT4 (BC-Bertini [28-31]and INCL [5-7]), c - MCNP6 (CEM03 [43-45] and INCL [5-7]); d - MARS15 [61-63].

| $E_p$, | Neutron production/part. | Neutron leakage/part. |
|---|---|---|

| GeV | a | b | | c | | d | a | b | | c | | d |
|---|---|---|---|---|---|---|---|---|---|---|---|---|
| | | BC-Bertini | INCL | CEM03 | INCL | | | BC-Bertini | INCL | CEM03 | INCL | |
| 4 | 375 | 412 | 393 | 467 | 439 | 386 | 78 | 79 | 75 | 88 | 66 | 67 |
| 2 | 197 | 215 | 176 | 220 | 217 | 221 | 46 | 46 | 37 | 52 | 42 | 40 |
| 1 | 89 | 95 | 85 | 100 | 96 | 101 | 24 | 24 | 21.7 | 27 | 24 | 21 |
| 0.66 | 51 | 50 | 43 | 57 | 61 | 55 | 16 | 15 | 13 | 17 | 15 | 14 |

Table 6. Number of fissions and neutron captures per particle in interaction of p beam with U-238 target with a radius of 60 cm and a length of 160 cm obtained using the following models: a - SHIELD [68, 69]; b - GEANT4 (BC-Bertini [28-31] and INCL [5-7]), c - MCNP6 (CEM03 [43-45] and INCL [5-7]); d - MARS15 [61-63].

| $E_p$, GeV | Number of fissions/part. | | | | | | Number of captures/part. | | | | | |
|---|---|---|---|---|---|---|---|---|---|---|---|---|
| | a | b | | c | | d | a | b | | c | | d |
| | | BC-Bertini | INCL | CEM03 | INCL | | | BC-Bertini | INCL | CEM03 | INCL | |
| 4 | 64 | 61 | 54 | 55 | 49 | 69 | 226 | 217 | 181 | 234 | 231 | 200 |
| 2 | 33 | 32 | 29 | 26 | 24 | 35 | 114 | 110 | 90 | 104 | 107 | 109 |
| 1 | 15 | 14.2 | 13.3 | 11 | 10 | 15 | 49 | 46 | 40 | 45 | 64 | 48 |
| 0.66 | 9.6 | 7.4 | 6.8 | 7 | 5 | 8 | 26 | 22 | 19 | 24 | 21 | 24 |

Table 7. Neutron production/loss in interaction of d beam with U-238 target with a radius of 15 cm and a length of 40 cm obtained using the following models: a - SHIELD [68, 69]; b – GEANT4 (BC-Bertini [28-31]); c - MCNP6 (CEM03 [43-45] and INCL [5-7]); d - MARS15 [61-63].

| $E_d$, | Neutron production/part. | Neutron leakage/part. |
|---|---|---|

| GeV/part. | a | b | c | | d | a | b | c | | d |
|---|---|---|---|---|---|---|---|---|---|---|
| | | | CEM03 | INCL | | | | CEM03 | INCL | |
| 4 | 542 | 569 | 590 | 537 | 671 | 334 | 320 | 346 | 316 | 382 |
| 382.42 | 311 | 300 | 315 | 313 | 367 | 191 | 168 | 186 | 184 | 148 |
| 1 | 163 | 163 | 180 | 167 | 194 | 100 | 92 | 106 | 99 | 111 |
| 0.66 | 111 | 110 | 123 | 110 | 133 | 68 | 62 | 74 | 66 | 77 |

Table 8. Number of fissions and neutron captures per particle in interaction of d beam with U-238 target with a radius of 15 cm and a length of 40 cm obtained using the following models: a - SHIELD[68, 69]; b – GEANT4 (BC-Bertini [28-31]); c - MCNP6 (CEM03 [43-45] and INCL [5-7]); d - MARS15 [61-63].

| $E_d$, GeV/part. | Number of fissions/part. | | | | | Number of captures/part. | | | |
|---|---|---|---|---|---|---|---|---|---|
| | a | b | c | | d | a | b | c | |
| | | | CEM03 | INCL | | | | CEM03 | INCL |
| 4 | 85 | 85,6 | 70 | 60 | 97.1 | 113 | 112,4 | 113 | 99 |
| 2 | 49 | 43,5 | 37 | 35 | 52.4 | 65 | 57,6 | 59 | 57 |
| 1 | 26 | 23,5 | 21 | 18 | 27.6 | 34 | 30,7 | 33 | 30 |
| 0.66 | 18 | 15,7 | 14 | 12 | 18.9 | 23 | 19,9 | 22 | 20 |

Table 9. Neutron production/loss in interaction of d beam with U-238 target with a radius of 30 cm and a length of 80 cm obtained using the following models: a - SHIELD [68, 69]; b – GEANT4 (BC-Bertini [28-31]); c - MCNP6 (CEM03 [43-45] and INCL [5-7]); d - MARS15 [61-63].

| $E_d$, | Neutron production/part. | Neutron leakage/part. |
|---|---|---|

| GeV/part. | a | b | c | | d | a | b | c | | d |
|---|---|---|---|---|---|---|---|---|---|---|
| | | | CEM 03 | INCL | | | | CEM 03 | INCL | |
| 4 | 736 | 799 | 825 | 818 | 830 | 233 | 235 | 232 | 230 | 274 |
| 2 | 396 | 392 | 429 | 426 | 438 | 131 | 116 | 129 | 130 | 150 |
| 1 | 206 | 204 | 231 | 216 | 231 | 70 | 64 | 72 | 71 | 82 |
| 0.66 | 129 | 134 | 155 | 138 | 153 | 45 | 44 | 52 | 48 | 55 |

Table 10. Number of fissions and neutron captures per particle in interaction of d beam with U-238 target with a radius of 30 cm and a length of 80 cm obtained using the following models: a - SHIELD [68, 69]; b – GEANT4 (BC-Bertini [28-31]); c - MCNP6 (CEM03 [43-45] and INCL [5-7]); d - MARS15 [61-63].

| $E_d$, GeV/part. | Number of fissions/part. | | | | | Number of captures/part. | | | |
|---|---|---|---|---|---|---|---|---|---|
| | a | b | c | | d | a | b | c | |
| | | | CEM 03 | INCL | | | | CEM 03 | INCL |
| 4 | 124 | 118 | 100 | 94 | 128.4 | 364 | 329 | 358 | 351 |
| 2 | 66 | 59 | 51 | 48 | 67 | 190 | 164 | 181 | 174 |
| 1 | 35 | 30 | 27 | 24 | 35 | 97 | 81 | 95 | 85 |
| 0.66 | 22 | 18 | 18 | 15 | 23 | 60 | 52 | 60 | 52 |

Table 11. Neutron production/loss in interaction of d beam with U-238 target with a radius of 60 cm and a length of 160 cm obtained using the following models: a - SHIELD [68, 69]; b – GEANT4 (BC-Bertini [28-31] and INCL [5-7]); c - MCNP6 (CEM03 [43-45] and INCL [5-7]); d - MARS15 [61-63].

| $E_d$, GeV/part. | Neutron production/part. | | | | | Neutron leakage/part. | | | | |
|---|---|---|---|---|---|---|---|---|---|---|
| | a | b | | c | | d | a | b | | c | | d |
| | | BC-Bertini | INCL | CEM03 | INCL | | | BC-Bertini | INCL | CEM03 | INCL | |
| 4 | 769 | 843 | 674 | 896 | 914 | 846 | 144 | 144 | 112 | 136 | 127 | 180 |
| 2 | 423 | 419 | 372 | 467 | 465 | 451 | 84 | 71 | 68 | 82 | 83 | 107 |
| 1 | 215 | 213 | 189 | 251 | 234 | 234 | 48 | 43 | 39 | 48 | 50 | 60 |
| 0.66 | 134 | 140 | 121 | 168 | 148 | 158 | 33 | 31 | 28 | 37 | 35 | 42 |

Table 12. Number of fissions and neutron captures per particle in interaction of d beam with U-238 target with a radius of 60 cm and a length of 160 cm obtained using the following models: a - SHIELD [68, 69]; b – GEANT4 (BC-Bertini [28-31] and INCL [5-7]); c - MCNP6 (CEM03 [43-45] and INCL [5-7]); d - MARS15 [61-63].

| $E_d$, GeV/part. | Number of fissions/part. | | | | | | Number of captures/part. | | | | |
|---|---|---|---|---|---|---|---|---|---|---|---|
| | a | b | | c | | d | a | b | | c | |
| | | BC-Bertini | INCL | CEM03 | INCL | | | BC-Bertini | INCL | CEM03 | INCL |
| 4 | 131 | 127 | 105 | 105 | 101 | 132 | 477 | 452 | 365 | 474 | 488 |
| 2 | 72 | 63 | 61 | 54 | 51 | 70 | 258 | 228 | 196 | 240 | 234 |
| 1 | 36 | 32 | 30 | 29 | 26 | 36 | 126 | 111 | 96 | 126 | 111 |
| 0.66 | 23 | 21 | 19 | 19 | 16 | 24 | 76 | 71 | 60 | 80 | 68 |

It can be seen that the difference between the predictions by different codes and models is, on average, within 30%, although, in certain cases it may be almost twice as high. This demonstrates the difficulties encountered in simulation of beam-matter interaction, especially in bulk targets and proves that in order to obtain reliable numerical picture of interaction, codes should be verified to experimental data for particular experimental conditions.

An important and least studied part of the neutron spectrum from large extended targets is that of fast neutrons. Below we give the numerical results obtained with different models on production and escape of fast neutrons for the target parameters corresponding to those of BURAN target: a radius of 60 cm and a length of 160 cm.

Table 13. Production and leakage of fast neutrons (En > 1 MeV) for U-238 target with a radius of 60 cm and a length of 160 cm irradiated by the proton beam obtained using the following models: a - SHIELD [68, 69]; b - GEANT4 (BC-Bertini [28-31] and INCL [5-7]), c - MCNP6 (CEM03 [43-45] and INCL [5-7]).

| $E_p$, GeV /part | Neutron production (En>1MeV)/part | | | Neutron leakage(En>1MeV)/part | | | | | Neutron leakage(En>1MeV)/ Neutron production (En>1MeV)( %) | | |
|---|---|---|---|---|---|---|---|---|---|---|---|
| | a | b | | a | b | | c | | a | b | |
| | | BC | INCL | | BC | INCL | CEM03 | INCL | | BC | INCL |
| 4 | 272.6 | 300.4 | 286.1 | 16.3 | 16.9 | 16.1 | 18 | 12 | 6 | 5.6 | 5.6 |
| 2 | 143.2 | 157.2 | 131.7 | 10.2 | 10.1 | 8.4 | 12 | 9 | 7.1 | 6.4 | 6.4 |
| 1 | 65.1 | 69.6 | 63.7 | 5.6 | 5.4 | 5.2 | 7 | 5 | 8.6 | 7.8 | 8.2 |
| 0.66 | 37.1 | 36.4 | 32.1 | 3.8 | 3.6 | 3.3 | 4 | 4 | 10.2 | 10 | 10.3 |

It can be seen from Table 13 that, on average, the agreement of the considered models is not bad. However, the discrepancy increases with increasing energy of the incident beam. Already for a proton energy of 1 GeV the model predictions may differ by as much as 40%, this discrepancy increasing to 43% for a 2 GeV proton beam and 50% for a 4 GeV proton beam. It should be noted that considered beam energies are of interest from the point of view of development of new concepts of nuclear power production with the aid of accelerated ion beams, which explains utmost importance of both theoretical and experimental study of these processes. Of course, neutron production and escape depends strongly on the target parameters: material and dimensions. Below we illustrate the effect of the target dimensions on these processes.

VI. Effect of target dimensions on neutron production and capture in a heavy bulk target irradiated by accelerated proton and deuteron beams

Irradiation of heavy extended targets by light ion beams is of substantial interest for development of the new concept of power production aided by an accelerator [70, 71] and for research toward transmutation of radioactive waste. Note that the neutron spectrum, especially the hard part, is of extreme importance in ADS nuclear power production. Although the fraction of fast neutrons is rather small, they carry a substantial part of the energy. Therefore, it is important that the geometry of the target is such that the produced fast neutrons do not leave its volume, carrying away a noticeable fraction of energy. Below, we estimate the appropriate target length and radius.

The numerical experiment on irradiation of bulk heavy targets by proton and deuteron beams is illustrated in Figs. 1 and 2. Experimental studies in this field are carried out at JINR. The comparison of calculations and measured data will be the topic of another paper.

The integrated number of fissions and captures per projectile, for the targets with a radius of 60 cm and different lengths is shown in Fig. 1, for 0.66 GeV proton, and 0.66, 1, 2 and 4 AGeV deuteron beams.

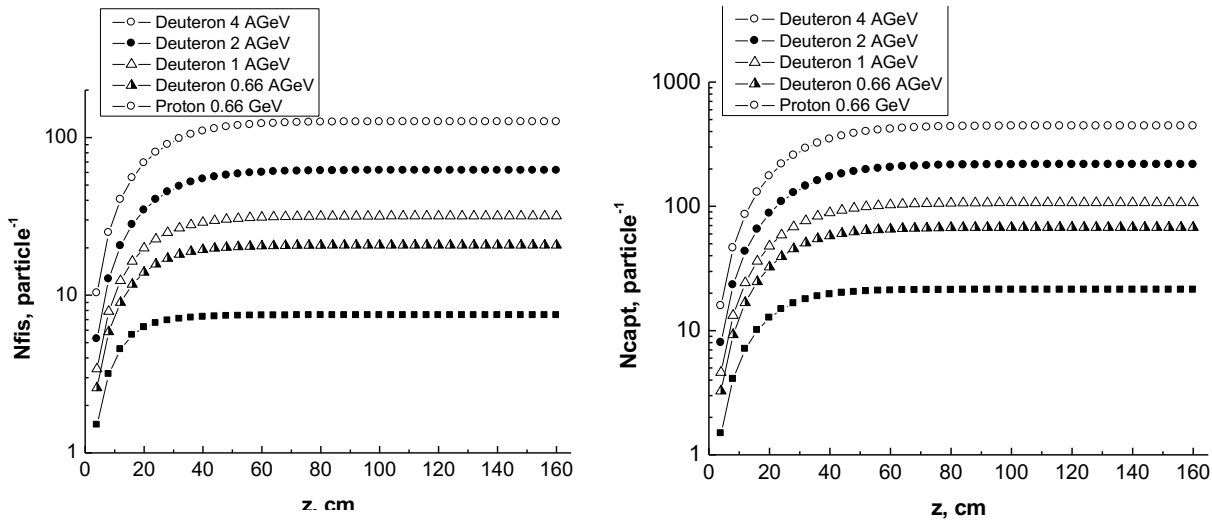

Fig. 1. Fissions and captures in the bulk $^{238}$U target depending on the target length.

It can be seen from Fig. 1 that both the number of fissions and the number of captures per one beam particle reach a plateau rather promptly, already for a target length of about 40 cm, for all considered beam types and energies.

The integrated number of fissions and captures per projectile, for the target with a length of 160 cm and different radii is shown in Fig. 2, for 0.66 GeV proton, and 0.66, 1, 2 and 4 AGeV deuteron beams.

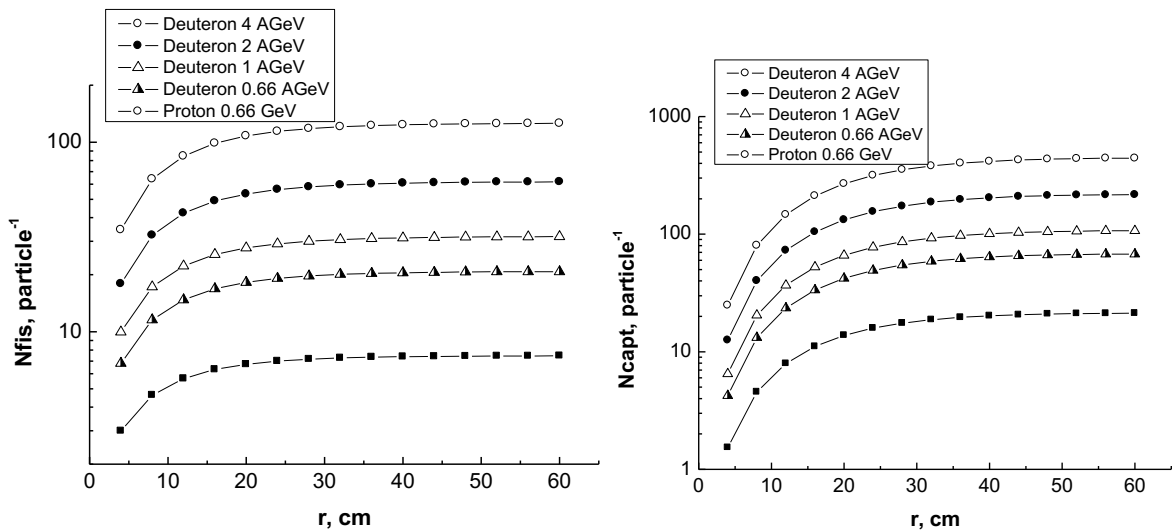

Fig. 2. Fissions and captures in the long irradiated target depending on the target radius.

It can be seen from Fig. 2 that, similar to the case of the target length variation, the number of fissions and neutron captures promptly reach a plateau with increasing target radius. The radius equal to 30 cm is roughly sufficient for it.

This testifies that a bulk target, being irradiated by proton and deuteron beams in a wide energy range, demonstrates a "saturation" mode when the number of fissions and captures per beam particle reach a stationary value and remain constant with further growth of the target dimensions.

CONCLUSIONS

The paper presented the comparative analysis of the basic cascade models of nuclear interactions applied in software for simulation of beam-target collisions: Liege intranuclear cascade, binary cascade, Bertini cascade, cascade exciton model and quark-gluon string model. The main physics approximations underlying these models were discussed.

The simulation of bulk heavy targets irradiated by accelerated proton and deuteron beams of energies from 0.66 GeV/nucleon to 4 GeV/nucleon was performed using five different software packages: SHIELD, GEANT4, MCNP6 and MARS15. The neutron production and escape from $^{238}$U targets of three sizes: a radius of 15 cm and a length of 40 cm, a radius of 30 cm and a length of 80 cm, and a radius of 60 cm and a length of 160 cm were analyzed. The beam and target parameters were chosen close to those studied experimentally at the accelerator complex of the Joint Institute for Nuclear Research.

It was shown that, on the whole, the agreement between the models and codes is within 30% and better, worsening in some cases to 50% discrepancy. The highest discrepancy was observed for the high energy part of the neutron spectrum, which is of importance in design of novel accelerator-aided nuclear power production facilities and nuclear waste transmutation issues. This indicates the need in further theoretical and experimental studies of inelastic interactions with production of fast neutrons and interaction of these neutrons with bulk heavy materials. The comparison of calculations and experimental data obtained at JINR will be the subject of another paper.